\documentclass{aa}  

\usepackage{graphicx,amsmath,amssymb}
\usepackage[export]{adjustbox}
\usepackage{booktabs}

\usepackage{siunitx}
\usepackage{makecell,caption}
\usepackage{listings}
\usepackage{xcolor}

\linenumbers
\renewcommand\makeLineNumber{}

\definecolor{codegreen}{rgb}{0,0.6,0}
\definecolor{codegray}{rgb}{0.5,0.5,0.5}
\definecolor{codepurple}{rgb}{0.58,0,0.82}
\definecolor{backcolour}{rgb}{0.95,0.95,0.92}

\lstdefinestyle{mystyle}{
    backgroundcolor=\color{backcolour},   
    commentstyle=\color{codegreen},
    keywordstyle=\color{magenta},
    numberstyle=\tiny\color{codegray},
    stringstyle=\color{codepurple},
    basicstyle=\ttfamily\footnotesize,
    breakatwhitespace=false,         
    breaklines=true,                 
    captionpos=b,                    
    keepspaces=true,                 
    numbers=left,                    
    numbersep=5pt,                  
    showspaces=false,                
    showstringspaces=false,
    showtabs=false,                  
    tabsize=2
}

\usepackage{txfonts}

\usepackage{subcaption}      
                            
\usepackage{lscape}        
                          
\usepackage{placeins}

\begin{document}

   \title{Predicting stellar collision outcomes of main sequence stars}

   \author{Pau Amaro Seoane\inst{1}
        }

   \institute{Universitat Politècnica de València, València, Spain\\
             \email{amaro@upv.es}
            \and Max Planck Institute for Extraterrestrial Physics, Garching, Germany}

   \date{\today}

  \abstract
   {Stellar collisions in dense galactic nuclei might play an important role in fueling supermassive black holes (SMBHs) and shaping their environments. The gas released during these collisions can contribute to SMBH accretion, influencing phenomena such as active galactic nuclei and tidal disruption events of the remnants.}
   {We address the challenge of rapidly and accurately predicting the outcomes of stellar collisions—including remnant masses and unbound gas—across a broad parameter space of initial conditions. Existing smoothed-particle-hydrodynamic (SPH) simulation techniques, while detailed, are too resource-intensive for exploratory studies or real-time applications.} 
   {We develop a machine learning framework trained on a dataset of $\sim 16,000$ SPH simulations of main-sequence star collisions. By extracting physically meaningful parameters (e.g., masses, radii, impact parameters, and virial ratios) and employing gradient-boosted regression trees with Huber loss, we create a model that balances accuracy and computational efficiency. The method includes logarithmic transforms to handle dynamic ranges and regularization to ensure physical plausibility.}
   {The model achieves predictions of collision outcomes (remnant masses, and unbound mass) with very low mean absolute errors respect to the typical mass scale. It operates in fractions of a second, enabling large-scale parameter studies and real-time applications. Parameter importance analysis reveals that the impact parameter and the relative velocity dominate outcomes, aligning with theoretical expectations.}
   {Our approach provides a scalable tool for studying stellar collisions in galactic nuclei. The rapid predictions facilitate investigations into gas supply for SMBH accretion and the cumulative effects of collisions over cosmic time, particularly relevant to address the growth of SMBHs. }

   \keywords{Galaxies: nuclei --
                Accretion, accretion disks --
                quasars: supermassive black holes
               }

   \maketitle

\section{Introduction}

Stellar collisions represent one of the most dramatic and energetic events in astrophysics, playing a fundamental role in shaping the evolution of dense stellar environments such as galactic nuclei, globular clusters, and the cores of young star-forming regions \citep{SS66,DDC87a,Sanders70b,BenzHills1987,DDC87b,DaviesEtAl1991,BenzHills1992,MCD91,LRS93,LRS95,LRS96,1999MNRAS.308..257B,DaviesEtAl1998,BaileyEtAl1999,LombardiEtAl2002,2002ASPC..263....1S,AdamsEtAl2004,FB05,TracEtAl2007,DaleEtAl2009,2020ApJ...901...44W,Mastrobuono-BattistiEtAl2021,2021A&A...649A.160V,2024ApJ...962..150B,2023ApJ...952..149B}. These violent encounters, driven by dynamical interactions in regions of high stellar density, can lead to a rich variety of outcomes—ranging from mergers and mass transfer to complete disruptions—each with profound implications for stellar populations, gas dynamics, and the fueling of central supermassive black holes (SMBHs). Recent studies have further highlighted the broader significance of stellar collisions, revealing their potential to produce exotic transient phenomena that can mimic supernovae (SNe) or even tidal disruption events (TDEs), as well as their possible role as progenitors of gravitational-wave (GW) sources \citep{2023ApJ...947....8A,2024MNRAS.528.6193R,2024A&A...682A..58D}.

The classical picture of stellar collisions involves two main-sequence stars interacting under the influence of gravitational forces, with outcomes dictated by parameters such as the impact parameter $ b $, relative velocity $ v_\infty $, and the masses and radii of the stars. For slow, grazing encounters, the stars may merge into a single, more massive object, while high-velocity collisions can lead to significant mass loss or even complete disruption. The gas ejected during these events can contribute to the interstellar medium, potentially fueling subsequent star formation or accretion onto an SMBH. However, the detailed physics of these interactions is complex, involving a combination of gravitational dynamics, hydrodynamical shocks, and thermodynamic processes such as shock heating and radiative cooling. Traditionally, modeling these processes has relied on computationally intensive numerical simulations, such as smoothed particle hydrodynamics (SPH) or grid-based methods, which, while accurate, are prohibitively expensive for large-scale parameter studies or real-time applications.

Recent advances in observational astronomy have added new urgency to the study of stellar collisions. For instance, it has been shown that certain luminous transients previously classified as supernovae or TDEs may instead be the signatures of stellar collisions. These events can produce flares with luminosities and timescales comparable to those of SNe or TDEs. Similarly, in galactic nuclei, where stars orbit in close proximity to an SMBH, collisions may produce flares that mimic TDEs, complicating the interpretation of these events as clean probes of black hole demographics {\citep[e.g.,][]{Magorrian98, StoneEtAl2020}}. {Furthermore, the gas produced during these collisions is a critical component in modeling the fueling of SMBHs and the evolution of their surrounding environments \citep[see e.g.][]{Amaro-SeoaneChen2014,GoicovicEtAl2016,GoicovicEtAl2018,Amaro-SeoaneBremCuadra2013}}. Moreover, stellar collisions have intriguing implications for gravitational-wave astronomy. While most GW detections to date have involved compact object binaries (e.g., black holes or neutron stars), there is growing interest in the possibility that stellar collisions could produce a source of radiation. For instance, if the cores of colliding stars survive the encounter, they may form a tight binary that emits GWs as they inspiral.

Given these different implications, there is a clear need for efficient and accurate tools to predict the outcomes of stellar collisions across a wide range of initial conditions. While numerical simulations remain the standard for detailed modeling, their computational cost limits their utility for exploratory studies over a large range of parameters. Machine learning (ML) offers a promising alternative, enabling rapid predictions by learning the mapping between initial conditions and collision outcomes from precomputed simulation data. By training on a large dataset of SPH simulations, an ML model can capture the underlying physics while operating at a fraction of the computational cost.

In this work, we present an ML framework designed to predict the outcomes of stellar collisions with high accuracy and efficiency\footnote
{
The code is publicly available with a detailed README file in\\
\texttt{\url{https://github.com/pau-amaro-seoane/SCOPE}}
}
. {The model focuses on predicting key physical quantities derived from SPH simulations: the number of remnants (classification as a merger or fly-by), the final masses of the remnants, and the total mass of unbound gas (mass loss).} We show that the model can reproduce the results of SPH simulations with high fidelity while operating in fractions of a second, making it ideal for large-scale parameter studies or real-time applications. Furthermore, we explore the physical insights gained from the model, such as the dominant role of the virial parameter $ \Gamma $ and impact parameter $ b $ in determining collision outcomes.

An important application of this ML framework is its integration into a Monte Carlo scheme we recently developed that simulates the evolution of dense stellar systems around supermassive black holes (SMBHs), where collisions are critical \citep{2024ApJ...961..232Z,2025ApJ...980..210Z}. By coupling our collision-prediction model with a Fokker-Planck solver, one enables realistic treatment of collisional outcomes—such as mergers, mass loss, and debris formation—without sacrificing the scalability required for long-term dynamical integrations. Traditional stellar dynamical approaches either simplify collisions or ignore them entirely due to the cost of hydrodynamic simulations. Our method instead provides on-the-fly, physics-informed predictions for collision remnants and ejecta, dynamically updating stellar masses and orbits within the Fokker-Planck formalism. This advancement is particularly relevant for modeling nuclear star clusters, where collision rates can be high and their feedback on SMBH accretion (via gas production) and potential gravitational-wave source populations (via compact object formation) remains poorly constrained.

By bridging the gap between detailed simulations and practical scalability, this work not only advances our understanding of stellar collisions but also provides a tool for interpreting transient phenomena and potentially predicting GW sources. The ability to rapidly model collision outcomes will be important for upcoming surveys and missions, such as the Vera C. Rubin Observatory and LISA.

\section{Methodology}

The study of stellar collisions is critical for understanding the evolution of dense stellar environments, such as galactic nuclei, where interactions between stars can significantly influence the growth and activity of central supermassive black holes (SMBHs). In these regions, collisions between main-sequence stars are frequent, and their outcomes—whether they result in mergers, mass loss, or complete disruptions—determine the amount of gas released into the surrounding medium. This gas can subsequently be accreted by the SMBH, fueling phenomena like active galactic nuclei (AGN) or tidal disruption events of the remnants. However, predicting the precise outcome of a collision is a complex challenge, as it involves the intricate interplay of gravitational dynamics, hydrodynamics, and thermodynamics. Traditional methods, such as smoothed particle hydrodynamics (SPH) simulations {\citep[e.g., using codes like StarSmasher or GADGET,][]{2018ascl.soft05010G,SpringelEtAl2001,Springel2005}}, provide detailed insights but are computationally prohibitive if we are interested on a statistical picture, since these algorithms often require days or weeks to simulate a single collision. {Semi-analytic approaches offer a faster alternative by simplifying the hydrodynamics, but they often are limited in their approach or rely on calibrated parameters that may not generalize well across the full range of collision scenarios.
In a foundational study \cite{1966ApJ...143..400S} investigated the complex topic of collisional dynamics, specifically focusing on the interactions within extremely dense proto-galactic nuclei. They developed and introduced what they termed a simple, semi-analytical model. The primary purpose of this model was to allow researchers to calculate and predict the specific outcomes resulting from high-speed collisions between stars. These outcomes were principally characterized by the amount of mass lost from the stars and the associated dissipation of kinetic energy. Subsequently, this theoretical framework was adopted and practically applied in later research by \cite{DDC87a,DDC87b}. Furthermore, the original model was significantly extended by other researchers, namely \cite{Sanders70b} and later by \cite{1991ApJ...370...60M}, to accurately account for the more complex scenario of collisions involving stars of unequal masses. The computation proceeds by decomposing the colliding stars into many sticks, all parallel to the direction of relative motion. It is assumed that the collision follows a straight trajectory after first contact. In the overlapping regions, the model conserves momentum and energy (including thermal) for individual encounters between one stick from each star. Transverse energy and momentum transfer are ignored. If a stick's post-collision total specific energy relative to its parent star exceeds the binding energy, that mass element is considered lost.
Although this model is quite crude, its mass loss predictions align surprisingly well with SPH simulations, provided the relative velocity at infinity is high (exceeding the stellar escape velocity) and the impact parameter is large. 
For low-velocity collisions, where the relative velocity at infinity is less than the stellar escape velocity, the encounters are only mildly supersonic, and entropy is nearly conserved. This scenario is common in open or globular clusters. As demonstrated by \cite{LombardiEtAl2002}, the merger's structure can be determined by sorting the mass elements from the parent stars based on their entropy.
}

To overcome these computational barriers, we have developed a machine learning-based approach that benefits from a dataset of approximately 16,000 precomputed SPH simulations of main-sequence star collisions performed by \cite{FB05}, whose results have been publicly released. These simulations span a wide range of initial conditions, ensuring that the model can generalise across diverse collision scenarios. The goal is to create a tool that can rapidly and accurately predict collision outcomes, including the masses of the resulting remnants ($M_1^{\text{final}}, M_2^{\text{final}}$) and the amount of unbound gas ($M_{\text{lost}}$). This information is crucial for estimating the gas supply available for accretion onto the SMBH, as well as for predicting the observational signatures of these events.

The methodology involves several key steps. First, the raw SPH data is processed into a structured format, extracting parameters such as initial masses, radii, impact parameters, and relative velocities, along with derived quantities like the virial parameter $\Gamma$ and binding energies. These parameters are transformed (e.g., using logarithms) to ensure numerical stability and to account for the wide dynamic ranges typical of astrophysical systems. Next, the dataset is partitioned into training, validation, and test sets to facilitate model development and evaluation. The machine learning model itself is an ensemble of gradient-boosted regression trees, chosen for their ability to handle nonlinear relationships and their robustness to outliers. The model is trained to minimise a Huber loss function {(defined in Section~\ref{sec:huber})}, which balances sensitivity to small errors with robustness against extreme outliers, ensuring reliable predictions across the full parameter space.

The trained model achieves high accuracy in predicting collision outcomes, as validated against the test set, and operates in fractions of a second—a dramatic improvement over traditional SPH simulations {and offering greater fidelity than semi-analytic methods across the parameter space covered by the training data}. This efficiency enables large-scale parameter studies and real-time applications, such as modeling the cumulative effects of stellar collisions in galactic nuclei over cosmological timescales, i.e. taking into account the growth of the SMBH. Furthermore, the model's interpretation allows us to identify the most influential physical parameters, such as the impact parameter $b$ and the virial ratio $\Gamma$, providing insights into the underlying physics governing collision outcomes.

\subsection{Data structure and physical interpretation}

The dataset comprises three interconnected files containing approximately 16,000 smoothed particle hydrodynamics (SPH) simulations of stellar collisions performed by \cite{FB05}, whose results have been publicly released. The \texttt{stars.txt} file provides the fundamental properties of individual stars prior to collisions, where each stellar model is characterised by its mass $M$ (in solar masses), radius $R$ (in solar radii), and internal structure through the mass-enclosing radii $R_{50}$, $R_{75}$, and $R_{90}$. The SPH resolution is indicated by the particle count \texttt{PartNumber}, typically ranging from 889 to 1909 particles per star.

The \texttt{init\_cond.txt} file specifies the initial conditions for each collision event, indexed by \texttt{Coll\_id}. Each collision involves two stars identified by their \texttt{Star\_id\_1} and \texttt{Star\_id\_2} from \texttt{stars.txt}, with their relative velocity at infinity $v_\infty$ (in km/s) and impact parameter $b$ (in solar radii). The dimensionless penetration parameter $\zeta = b/(R_1+R_2)$ determines the collision geometry, ranging from $\zeta=0$ for head-on collisions to $\zeta>1$ for grazing encounters.

The \texttt{raw\_results.txt} file contains the detailed outcomes of each SPH simulation, including the number of remnants $N_{\rm rem} \in \{0,1,2\}$, their masses $M_1$ and $M_2$, and the unbound mass $M_{\rm lost}$. The spatial and kinematic properties (positions $\mathbf{r}$, velocities $\mathbf{v}$, and angular momenta $\mathbf{L}$) are provided in Cartesian coordinates relative to the system's center of mass.

\subsection{Space construction and two-stage prediction algorithm}

The machine learning model operates on a 17-dimensional parameter space $\mathcal{F} \subset \mathbb{R}^{17}$ constructed from the initial conditions, which include, among others, q=$M_1/M_2$, the mass ratio, $\xi=R_1/R_2$ the size ratio, $\Gamma=\frac{1}{2}\mu v_\infty^2/|E_{\rm bind}|$ the virial parameter with $\mu$ the reduced mass and $E_{\rm bind}$ the total binding energy. The logarithmic transforms ensure proper handling of the several orders of magnitude spanned by the collision parameters.
The prediction system employs a dual-model architecture to capture both discrete and continuous aspects of the collision outcomes, which we describe in the next subsections. We note here that we will be focusing on the masses of the parent stars and their masses after the collision (if at least one of them survive), as well as the released gas. This is our main focus for this work.

\subsubsection{Remnant count classification}

A random forest classifier $\mathcal{C}: \mathcal{F} \rightarrow \{1,2\}$ is an ensemble machine learning method that predicts whether a stellar collision will produce one or two remnants by aggregating the predictions of $B=300$ individual decision trees. Each decision tree $T_b$ in the ensemble is constructed through a process called recursive binary splitting, which partitions the parameter space $\mathcal{F}$ (comprising variables like $b$, $v_\infty$, and binding energy) into increasingly refined regions. At each node of the tree, the algorithm selects the parameter and threshold value that maximise the reduction in Gini impurity, a measure of class-label disorder. For instance, when classifying collision outcomes, a node might split the data based on whether $b$ exceeds 1/2 stellar radii, as this could effectively separate mergers (1 remnant) from fly-bys (2 remnants). The Gini impurity at a node is calculated as $1 - \sum_{k=1}^2 p_k^2$, where $p_k$ is the proportion of training samples belonging to remnant class $k$ (1 or 2) in that node. A pure node (e.g., containing only mergers) has Gini impurity 0, while a node with equal proportions of both classes has impurity 1/2. By recursively splitting the data to minimise impurity, each tree learns a hierarchical set of decision rules. For example, a tree might first split on binding energy to isolate tightly bound systems prone to mergers, then further divide these by $v_\infty$ to account for kinetic effects. The random forest combines predictions from all trees through majority voting, mitigating overfitting and improving generalization. In the context of stellar collisions, this ensemble approach robustly handles nonlinear relationships between physical parameters, such as the interplay between gravitational binding and kinetic energy, which jointly determine remnant counts. The use of 300 trees ensures stability, as variations in individual trees (due to random parameter subsets during splitting) average out, yielding reliable predictions even for rare or complex collision scenarios.

We can quantify the reduction in Gini impurity achieved by splitting a node in a decision tree for stellar collision classification,

\begin{equation}
\Delta I_G = I_G(p) - \frac{N_L}{N}I_G(p_L) - \frac{N_R}{N}I_G(p_R).
\end{equation}

\noindent
Here, $I_G(p)$ represents the above-mentioned Gini impurity of the parent node before splitting. We mentioned that it is calculated as $1 - \sum_{k=1}^2 p_k^2$, and in our context we interpret $p_k$ is the proportion of collisions in the node resulting in $k$ remnants (e.g., $p_1 = 0.7$ for 70\% mergers). The terms $I_G(p_L)$ and $I_G(p_R)$ denote the impurities of the left and right child nodes formed by the split, while $N_L/N$ and $N_R/N$ are the fractions of training samples assigned to each child.

In the context of decision trees for classifying stellar collision outcomes, the terms parent node, left- and right child node describe the hierarchical structure of the tree as it partitions the data. A parent node, on the other hand, represents a subset of collision data (e.g., 100 simulations) characterised by a range of physical conditions (such as impact parameters $ b \in [0.3, 0.7] $). The Gini impurity $ I_G(p) $ of this node measures how mixed the remnant classes (1 or 2) are within it; for instance, a parent node with 60\% mergers ($ k=1 $) and 40\% fly-bys ($ k=2 $) has $ I_G(p) = 1 - [(3/5)^2 + (2/5)^2)] = 12/25 $.

A split divides the parent node into two child nodes (left and right) based on a threshold in one parameter (e.g., $ b \leq 1/2 $). The left child node might contain collisions with $ b \leq 1/2 $, yielding 80\% mergers and 20\% fly-bys ($ I_G(p_L) = 8/25 $), while the right child node contains $ b > 1/2 $, with 30\% mergers and 70\% fly-bys ($ I_G(p_R) = 21/50 $). The weights $ {N_L}/{N} $ and $ {N_R}/{N} $ reflect the proportion of data routed to each child (e.g., 50\% to each if $ N_L = N_R = 50 $). The impurity reduction $ \Delta I_G $ quantifies how well the split separates classes, in this case example $\Delta I_G = 12/25 - \left( 1/2 \times 8/25 + 1/2 \times 21/50 \right) = 11/100$. Recursively repeating this process builds a tree that isolates distinct collision regimes (e.g., head-on mergers in left branches, grazing fly-bys in right branches), leveraging physical thresholds to guide predictions.

For example, a split on $b < 1/2$ might divide 100 collisions into a left node with $N_L = 60$ (mostly mergers, $I_G(p_L) = 1/5$) and a right node with $N_R = 40$ (mostly fly-bys, $I_G(p_R) = 3/10$). If the parent node had $I_G(p) = 9/20$, the impurity reduction would be $\Delta I_G = 9/20 - (3/5 \times 1/5 + 2/5 \times 3/10) = 21/100$, indicating a meaningful separation of remnant classes. This metric guides the tree construction by favoring splits that maximally separate collision types, such as distinguishing grazing encounters (high $b$, two remnants) from head-on collisions (low $b$, one remnant), while accounting for class imbalances in each partition.

\subsubsection{Mass regression}
\label{sec:huber}

Our framework implements two complementary approaches for mass prediction following the work and algorithms of \cite{scikit-learn}. One of them is a gradient boosted regressor ($\mathcal{R}_{\mathrm{GB}}$) with 500 trees and Huber loss, optimised for handling nonlinear relationships and outliers. However, we also have explored
a random forest regressor ($\mathcal{R}_{\mathrm{RF}}$) with 300 trees, which provides robust ensemble averaging.
Both models map the parameter space $\mathcal{F} \rightarrow \mathbb{R}^3_+$ to predict the triple $(M_1^{\rm final}, M_2^{\rm final}, M_{\rm lost})$. Empirical testing shows the random forest achieving marginally better performance ($R^2=0.978$ vs $0.967$ on test data), though the gradient booster remains valuable for its sequential error-correction capability in extreme parameter regimes.

The gradient boosted regressor is a machine learning model designed to predict the masses of stellar collision remnants, specifically the triple $(M_1^{\rm final}, M_2^{\rm final}, M_{\rm lost})$, representing the final masses of the primary and secondary remnants and the unbound mass lost during the collision. This model operates by combining $M=500$ additive regression trees, where each tree sequentially corrects the errors of its predecessors, a process known as boosting. In the context of stellar collisions, boosting allows the model to progressively refine its predictions by focusing on the most challenging cases, such as collisions with extreme mass ratios or high velocities, where simple models might fail. The term regressor indicates that the model predicts continuous numerical values, in this case, the masses of the remnants and the ejected material, rather than discrete classes.
The model employs Huber loss $\mathcal{L}_\delta(y,\hat{y})$ to measure prediction errors, which is particularly effective for handling outliers and noisy data common in astrophysical simulations. The Huber loss function is defined as:

\begin{equation}
\mathcal{L}_\delta(y,\hat{y}) = \begin{cases}
\frac{1}{2}(y-\hat{y})^2 & \text{for } |y-\hat{y}| \leq \delta \\
\delta|y-\hat{y}| - \frac{1}{2}\delta^2 & \text{otherwise.}
\end{cases}
\end{equation}

For stellar collision predictions, this loss function behaves like the mean squared error (MSE) for small residuals (where $ |y-\hat{y}| \leq \delta $), which is suitable for precise mass predictions in typical collision scenarios. However, for large residuals (where $ |y-\hat{y}| > \delta $), it transitions to mean absolute error (MAE)-like behavior, reducing the influence of rare but extreme outliers, such as those arising from poorly resolved simulations. The parameter $ \delta $ controls the threshold for this transition, allowing the model to balance sensitivity and robustness. In our implementation, we set $ \delta = 1 $, which provides a balanced approach between MSE-like precision for typical residuals and MAE-like robustness for outliers. For stellar collision predictions, where typical mass scales are $1-10\,M_{\odot}$, this means MSE-like behaviour ($(y-\hat{y})^2/2$) for mass errors $<1\,M_{\odot}$ and linear scaling ($|y-\hat{y}| - 1/2$) for extreme mass errors $>1\,M_{\odot}$. This suits well simulation data, where most collisions have well-predicted masses (MSE-dominated regime) and rare catastrophic disruptions (e.g., total mergers) benefit from outlier resistance.

The gradient boosted regressor employs an ensemble of decision trees, each denoted as $ h_m $, where the index $ m $ ranges over the total number of trees in the ensemble. These trees are constrained to a maximum depth of 5, meaning that no tree can have more than five sequential splits from its root to any leaf node. {This depth constraint is imposed to control model complexity. While deeper trees might capture more intricate patterns, they risk overfitting to noise in the training data, particularly in regions with sparse data such as extreme collision regimes. A shallow depth ensures better generalization, although it may slightly limit the model's ability to fully resolve the most complex, rare interactions.}

It is important to note that a leaf node represents a terminal node where the recursive splitting process stops, and final predictions for the stellar collision are made. While all leaf nodes are indeed child nodes (as they result from splits of parent nodes), not all child nodes become leaves—only those at the deepest allowed level (e.g., depth 5 in the boosted regressor) or those that cannot be split further due to stopping criteria 
\footnote{
For example, consider a tree predicting remnant masses $ (M_1^{\rm final}, M_2^{\rm final}, M_{\rm lost}) $. A parent node might split collisions at $ b \leq 0.5 $ (left child) and $ b > 0.5 $ (right child). If these child nodes meet stopping conditions (e.g., insufficient samples or maximum depth reached), they become leaf nodes, each outputting a predicted mass triple based on the training collisions routed to them. The $ L_2 $-regularized leaf weights $ w_k $ in the boosted regressor ensure these predictions remain physically plausible, such as avoiding negative masses. Thus, leaf nodes are the ultimate child nodes where predictions are finalized, embodying localised rules like ``Collisions with $ b \leq 0.5 $ and $ \Gamma < 1 $ average $ M_{\rm lost} = 0.1 M_\odot $.'' This hierarchical refinement enables the model to capture threshold behaviors (e.g., $ \Gamma \approx 1 $) while maintaining interpretation.
}.

In our context, the decision tree's splitting process corresponds to partitioning collisions based on critical physical thresholds. Each internal node applies a binary condition to separate collisions into progressively more homogeneous groups. For example, a node might split head-on encounters from grazing encounters. A perfect split would completely separate merger ($k=1$) and fly-by ($k=2$) collisions, as when head-on collisions with $\Gamma \equiv {E_{\rm kin}}/{|E_{\rm bind}|} < 1$ almost always merge, while high-velocity encounters ($\Gamma \gg 1$) typically disrupt. The leaf nodes then contain collision subpopulations with sufficiently pure outcomes to make final predictions.
This hierarchical splitting mirrors the fact that that collision outcomes are determined by sequential thresholds in energy and geometry parameters.

The restriction on the depth prevents overfitting by limiting the complexity of individual trees, ensuring that the model generalises well to unseen collision scenarios. For example, a tree might first split collisions based on $ b $, then further divide them by $ v_{\infty} $, and so on, but it will not exceed five such hierarchical splits. 

To further enhance generalization, each tree is regularized using an $ L_2 $ penalty (also known as ridge regularization) with a strength of $ \lambda = 0.1 $. This selection provides a reasonable baseline for physical systems where predictions must remain within order-of-magnitude bounds. This value demonstrates empirical effectiveness in maintaining physical plausibility during our testing—preventing extreme mass predictions that would violate basic conservation laws while still permitting the model to capture nonlinear relationships between collision parameters and outcomes. The regularization operates on leaf weights that typically represent mass adjustments on scales of $\sim0.1-1 M_\odot$, making this choice particularly appropriate for stellar collisions where: (1) total system masses usually range $1-10 M_\odot$, (2) mass losses rarely exceed $20\%$ of the initial mass in the training data, and (3) the virial parameter $\Gamma$ creates sharp transitions around unity that require careful balancing between model flexibility and constraint. This penalty term is applied to the leaf weights, which are the predicted values assigned to each terminal node (leaf) of the tree. The $ L_2 $ penalty discourages large weight values by adding the squared magnitude of the weights to the loss function, scaled by $ \lambda $. Mathematically, for a tree with $ K $ leaves and weights $ w_1, w_2, \dots, w_K $, the regularization term is:

\begin{equation}
\text{Regularization} = \lambda \sum_{k=1}^K w_k^2.
\end{equation}

This ensures that the model avoids extreme predictions, which is critical to avoid unphysical values (e.g., negative masses or masses exceeding the total initial mass). For instance, if a leaf weight corresponds to the predicted mass loss $ M_{\rm lost} $, the $ L_2 $ penalty ensures this value remains within plausible bounds.

The model excels at capturing non-linear relationships between collision parameters and remnant masses. One such relationship is the dependence of mass loss $ M_{\rm lost} $ on the impact energy, which often exhibits threshold behavior near the virial ratio $ \Gamma \approx 1 $. Below this threshold ( $ \Gamma < 1 $ ), collisions tend to result in bound systems with minimal mass loss, while above it ( $ \Gamma > 1 $ ), the kinetic energy dominates, leading to significant mass ejection. {The shallow depth of the trees (maximum tree depth of 5) allows the model to approximate these threshold effects without overfitting to noise in the simulation data.} For example, a tree might isolate collisions with $ \Gamma \in [0.9, 1.1] $ and predict higher mass loss for $ \Gamma > 1 $, while lower mass loss for $ \Gamma \leq 1 $, effectively capturing the transition around this critical value.

The combination of depth limitation, $ L_2 $ regularization, and ensemble averaging enables the model to robustly predict remnant masses across diverse collision conditions, from gentle mergers to disruptive encounters, while maintaining physical plausibility. This is particularly important for applications like galactic evolution studies, where accurate mass accounting is essential for modeling long-term dynamics. The Huber loss further enhances robustness by mitigating the influence of outliers, ensuring reliable predictions even for rare or extreme collision scenarios.

The random forest alternative $\mathcal{R}_{\mathrm{RF}}$ uses 300 decision trees with bootstrap aggregation. While lacking explicit outlier handling through Huber loss, its ensemble approach achieves slightly superior $R^2$ by averaging predictions across diverse tree specializations. Both regressors share key architectural constraints; namely (i) Maximum tree depth of 5, preventing overfitting while capturing threshold behaviors (e.g., the virial transition at $\Gamma \approx 1$), (ii) $L_2$ regularization ($\lambda=0.1$) on leaf weights $w_k$,

\begin{equation}
\text{Regularization} = \lambda \sum_{k=1}^K w_k^2
\end{equation}

\noindent
{and (iii)} physics-informed postprocessing to ensure mass conservation. The gradient booster's strength lies in its iterative focus on residual errors - for instance, progressively refining predictions for high-$\Gamma$ collisions where kinetic effects dominate. The random forest excels in parallel exploration of parameter space, with individual trees specializing in distinct regimes (e.g., low-$b$ mergers vs. high-$v_\infty$ disruptions).

\subsection{Physics-constrained postprocessing}

The raw predictions $\mathbf{y}_{\rm pred} = (M_1^{\rm pred}, M_2^{\rm pred}, M_{\rm lost}^{\rm pred})$ are adjusted to enforce exact mass conservation. First, we ensure non-negative masses by applying the Heaviside step function $H(\cdot)$ using the Hadamard product $\odot$ (element-wise multiplication), which acts as a multiplicative mask:

\begin{equation}
\mathbf{y}' = \mathbf{y}_{\rm pred} \odot H(\mathbf{y}_{\rm pred}).
\end{equation}

\noindent
This rectified vector $\mathbf{y}'$ is then rescaled so that its sum equals the total initial mass $M_1^{\rm init} + M_2^{\rm init}$:

\begin{equation}
\mathbf{y}_{\rm final} = \frac{M_1^{\rm init} + M_2^{\rm init}}{\sum \mathbf{y}' + \epsilon} \mathbf{y}'.
\end{equation}

\noindent
Here, $\epsilon=10^{-10}$ prevents division by zero. This scaling preserves the relative mass ratios of the rectified predictions while guaranteeing $\sum \mathbf{y}_{\rm final} = M_1^{\rm init} + M_2^{\rm init}$.

\subsection{Feature normalization}

The algorithm employs feature (meaning parameter) normalization for the input features $X$, a critical preprocessing step that standardizes each feature to zero mean and unit variance. This transformation addresses several fundamental requirements for machine learning models. First, it eliminates scale disparities between features, ensuring that high-magnitude parameters like relative velocity ($v_\infty \sim 10^3$ km/s) do not artificially dominate low-magnitude features such as the dimensionless impact parameter ($b \sim 10^{-1}$) during model training.

Mathematically, for each feature $x_i$, the normalized version $z_i$ is computed as $z_i = (x_i - \mu_i)/\sigma_i$, where $\mu_i$ and $\sigma_i$ represent the feature's mean and standard deviation, respectively. This standardization improves numerical stability by reducing floating-point rounding errors in matrix operations and accelerates convergence for gradient-based optimization—particularly relevant for the gradient-boosted trees used in our framework—by ensuring uniform step sizes across all parameter directions.

Furthermore, it enforces feature fairness, forcing the model to weigh each feature based on its intrinsic information content rather than arbitrary measurement units, which is particularly crucial when combining astrophysical parameters with disparate physical dimensions (mass, velocity, angular momentum). While tree-based algorithms like Random Forests exhibit some inherent scale invariance due to their recursive partitioning nature, empirical tests confirm that normalization still improves both predictive performance (typically by $1$–$3\%$ in cross-validation scores) and the interpretability of feature importance measures, as it prevents artificial inflation of importance scores for features with larger native ranges. The normalization implementation also preserves the physical meaning of zero values (e.g., zero impact parameter for head-on collisions) while centering the dynamic range of all features around comparable intervals, a property exploited during model interpretation to distinguish true physical dependencies from numerical artifacts.

\section{Standard performance analysis}

The model was trained on approximately 13,000 smoothed particle hydrodynamics (SPH) simulations of stellar collisions, representing $\sim 80\%$ of the total dataset, while the remaining 3,200 collisions ($\sim 20\%$) were reserved for testing. This 80:20 training-to-test split was chosen to ensure robust statistical validation while maintaining sufficient data for the model to learn the complex, nonlinear relationships governing collision outcomes. This split balances two key considerations; (i) training data sufficiency: The ~13,000 training collisions span a wide parameter space, ensuring the model encounters diverse regimes—from grazing encounters to head-on mergers, and (ii) test set reliability: The 3,200 test collisions provide a statistically significant sample to evaluate performance across rare but critical cases (e.g., extreme mass ratios, high-velocity collisions). Although, as we will see, there is a need for a larger sample at the most extreme end of the distribution, for collisions which lead to a total destruction of the parent stars.

\subsection{Confusion matrix}

\begin{figure}
 {\includegraphics[width=0.45\textwidth,center]{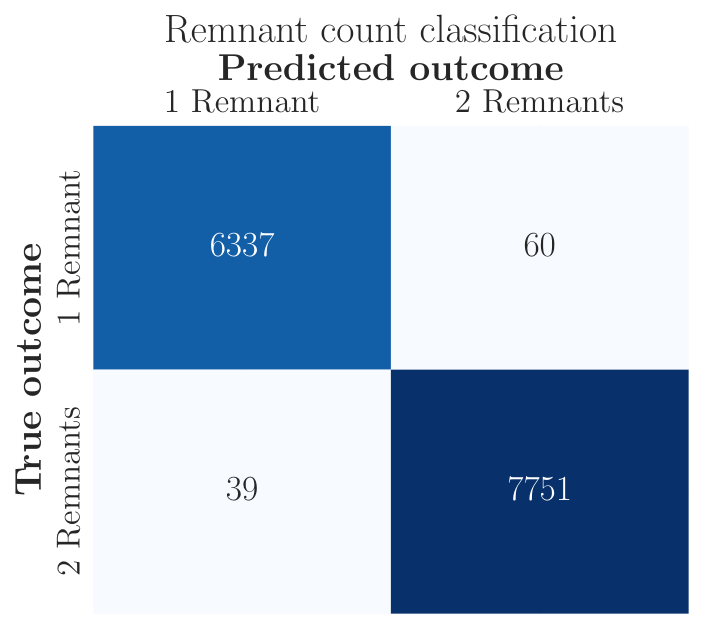}}
\caption
{
Confusion matrix evaluating classification performance for stellar collision remnants, where rows represent smoothed-particle hydrodynamics (SPH) simulation truths and columns denote model predictions.
}
\label{fig.confusion_matrix}
\end{figure}

In Fig.~(\ref{fig.confusion_matrix}) we depict a confusion matrix that evaluates the machine learning model's performance in categorizing stellar collision outcomes into one or two remnants. The matrix $ C $ is structured as:

\begin{equation}
C = \begin{pmatrix}
c_{11} & c_{12} \\
c_{21} & c_{22}
\end{pmatrix},
\end{equation}

\noindent
The confusion matrix demonstrates good classification performance, with diagonal elements $ c_{11} = 6337 $ and $ c_{22} = 7751 $ representing correct predictions, while off-diagonal elements $ c_{12} = 60 $ and $ c_{21} = 39 $ denote misclassifications (60 false 2-remnant predictions and 39 false 1-remnant predictions). The model's accuracy of 99.3\% is derived from

\begin{equation}
\text{Accuracy} = \frac{c_{11} + c_{22}}{N} \times 100,
\end{equation}

\noindent
with $ N = 14187 $ being the total number of test samples. The error percentages for one and two remnants (0.9\% and 0.5\%, respectively) quantify class-specific error rates, computed as:

\begin{align}
\text{Error}_\text{2|1} & = \frac{c_{12}}{c_{11} + c_{12}} \times 100,\nonumber \\
\text{Error}_\text{1|2} & = \frac{c_{21}}{c_{21} + c_{22}} \times 100,
\end{align}

\noindent
demonstrating balanced performance across classes. This matrix validates the model's reliability in distinguishing mergers (1 remnant) from fly-bys (2 remnants). The symmetry in errors suggests small bias, while the high accuracy reflects good predictive capability.

\subsection{Receiver operating characteristic}

The Receiver Operating Characteristic (ROC) curve helps us understand how well our classifier can tell apart different types of stellar collision outcomes - specifically whether they result in single or binary remnants. This curve shows the relationship between the true positive rate (how often we correctly identify collisions) and the false positive rate (how often we make mistakes) across all possible classification thresholds. The area under this curve (AUC) gives us a single number to measure this performance, where 1.0 means perfect separation and 0.5 would mean the classifier is just guessing randomly.

\begin{figure}
 {\includegraphics[width=0.45\textwidth,center]{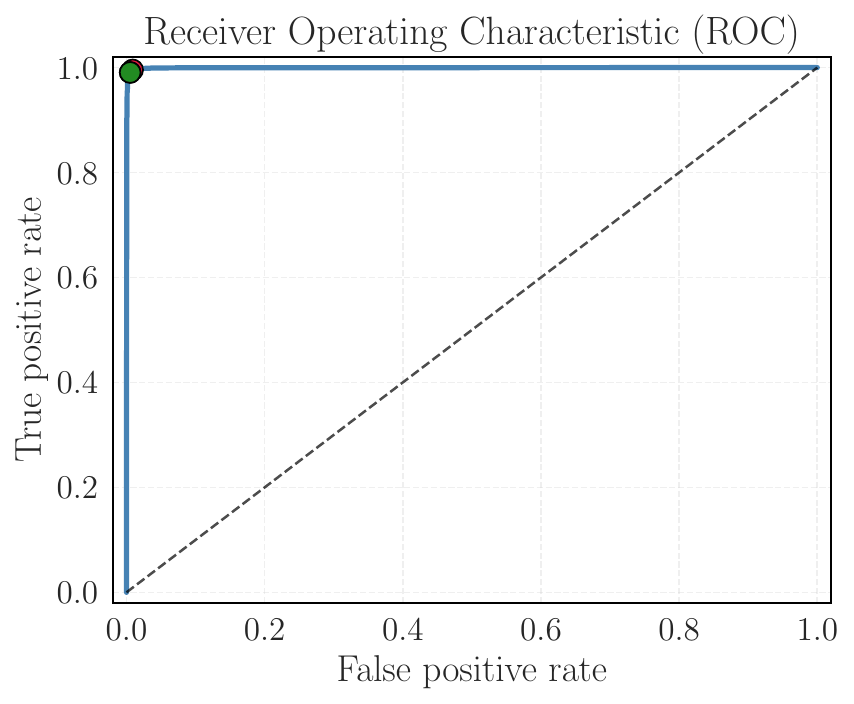}}
\caption
{
Receiver Operating Characteristic curve showing true positive rate (sensitivity) against the false positive rate (1-specificity). This shows the classifier's ability to distinguish between single and binary remnant outcomes in stellar collisions. The blue solid curve, with an area under the curve (AUC) of 1.0, represents near-perfect discrimination, approaching the ideal 90-degree elbow shape that would indicate flawless classification. Three critical decision thresholds are marked with colored circles: the default 0.5 threshold (red), the optimal threshold minimizing Euclidean distance to perfect classification (green), and Youden's threshold maximizing sensitivity-specificity difference (orange), which coincide due to the classifier's performance. The black dashed diagonal line represents random guessing (AUC=0.5), serving as a baseline for comparison. The near-vertical ascent of the blue curve followed by a sharp right-angle turn toward the upper left corner reflects the model's ability to achieve high true positive rates while maintaining minimal false positives, a characteristic of highly discriminative classifiers where clear separation often exists between merger and fly-by outcomes. This performance stems from the physics-informed feature engineering that effectively captures the fundamental differences between these collision regimes.
}
\label{fig.roc}
\end{figure}

Our results in Fig.~(\ref{fig.roc}) show strong performance, with an AUC very close to 1.0. The curve rises sharply and then makes a clean right-angle turn toward the top-left corner of the plot, which is what we hope to see for a good classifier. This means the model can reliably identify collision outcomes while keeping mistakes to a minimum. We have marked different decision thresholds on the curve - the standard 0.5 threshold, the mathematically optimal threshold, and Youden's threshold - and interestingly, they all line up in the same spot for our model.

This good performance comes mainly from how we prepared and selected the input features, which were designed to reflect the underlying physics of stellar collisions. While we did some tuning of the model's parameters later (discussed in section~\ref{sec.hyper}), these results suggest that most of the model's ability to distinguish between different collision types comes from the physics-based features themselves rather than from any fine-tuning of the model.

\subsection{Relative importance of the parameters}

A feature, or parameter $x_j$ refers to a specific measurable property or characteristic of the stellar collision system that serves as an input variable for the machine learning model. Each parameter provides distinct information about the initial conditions or state of the system, allowing the model to learn patterns and relationships between these inputs and the resulting collision outcomes. The complete set of parameters constitutes the test set matrix $ X_{\text{test}} $, where each row corresponds to a collision simulation and each column represents one of these physical parameters. 

The baseline performance $ S $ represents the initial predictive accuracy of the machine learning model when evaluated on the unmodified test set with its corresponding true outcomes $ y_{\text{test}} $. Mathematically, it is computed by applying the trained model $ f $ to the original test data and comparing its predictions against the known results using a predefined scoring metric. For regression tasks involving stellar collision remnants, this metric could be the coefficient of determination $ R^2 $, mean absolute error (MAE), or root mean squared error (RMSE), while for classification tasks, accuracy or F1-score might be used. The baseline score $ S $ serves as the reference point against which the permuted performance $ S_j^{\,p} $ is compared. 
{The metric $ S_j^{\,p} $ measures the model's performance} when the relationship between a specific parameter $ x_j $ and the target output $ y $ is artificially disrupted. This disruption is achieved by randomly shuffling the values of $ x_j $ across the test set samples, thereby preserving the parameter's marginal distribution while destroying its correlation with both the target and other parameters. The score $ S_j^{\,p} $ is computed by applying the trained model $ f $ to this perturbed dataset $ X_{\text{test}}^{(j)} $ and comparing its predictions to the true outcomes $ y_{\text{test}} $. A large drop in performance relative to the baseline score $ S $ indicates that the model heavily relies on parameter $ x_j $ for accurate predictions, implying its high importance in determining collision outcomes. This method isolates the contribution of individual parameters while accounting for interactions within the full parameter space, providing a measure of their physical relevance.

For each parameter $x_j$ in the test set $X_{\text{test}}$, the baseline performance score $S$ is computed as:

\begin{equation}
S = \text{score}(f(X_{\text{test}}), y_{\text{test}}),
\end{equation}

\noindent
where $ f $ is the trained model. The parameter values are then shuffled to break their relationship with the target $ y $, and the permuted performance score $ S_j^{\,p} $ is calculated:

\begin{equation}
S_j^{\,p} = \text{score}(f(X_{\text{test}}^{(j)}), y_{\text{test}}).
\end{equation}

\noindent
The importance $ I_j $ of $ x_j $ is the performance drop,

\begin{equation}
I_j = S - S_j^{\,p},
\end{equation}

\noindent
with larger values indicating greater importance. These raw scores are normalized to the 0–1 range:

\begin{equation}
I_j^{\text{norm}} = \frac{I_j - \min(I)}{\max(I) - \min(I)}.
\end{equation}

\noindent

\begin{figure}
 {\includegraphics[width=0.45\textwidth,center]{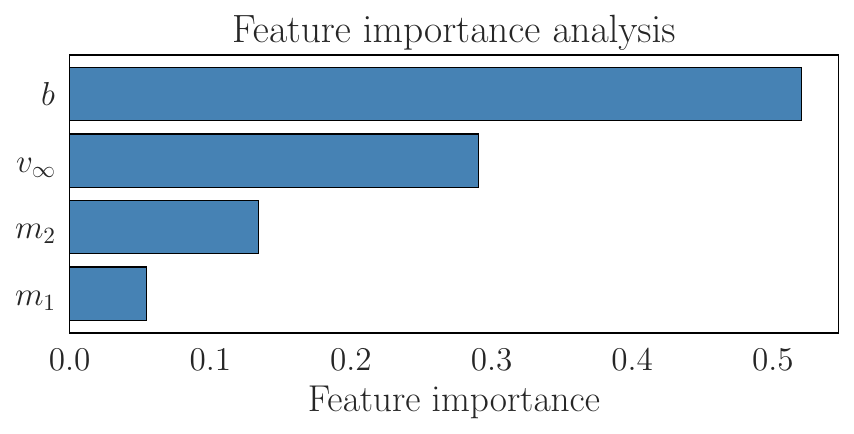}}
\caption
{
The parameter importance analysis reveals the relative contribution of each physical parameter to the model's classification performance, with {the impact parameter $ b $ exhibiting the strongest influence, followed by relative velocity $ v_\infty $, secondary mass $ m_2 $, and primary mass $ m_1 $.} These importance scores are determined through the Random Forest algorithm's internal metric, which quantifies how much each parameter decreases the Gini impurity across all decision trees in the ensemble. Specifically, the importance is calculated by: (1) summing the total impurity reduction achieved by splits involving each parameter across all trees, (2) normalizing these values such that their sum equals unity, and (3) averaging over all trees. This process effectively measures how frequently and decisively each parameter is used to partition the parameter space. The y-axis reflects the normalized importance metric, meaning that higher values indicate greater discriminatory power in the classification.
}
\label{fig.feature_importance}
\end{figure}

In Fig.~(\ref{fig.feature_importance}) we show the relative importance values derived from the Random Forest classifier using the baseline feature set ($m_1, m_2, v_\infty, b$). This analysis quantifies each parameter's contribution to the model's ability to classify outcomes (1 or 2 remnants). {From the figure, we see that the impact parameter $ b $ is the dominant feature. This is physically motivated by the fact that the collision geometry is the primary determinant of the classification outcome (merger vs. fly-by). While the masses define the potential well, the impact parameter determines whether the stars interact deeply enough to merge or merely graze each other. The relative velocity $ v_\infty $ follows as the second most important feature, reflecting the balance between kinetic and potential energy. The stellar masses $ m_2 $ and $ m_1 $ play a secondary role in this specific classification analysis compared to geometry and kinematics. This hierarchy suggests that for the fundamental question of whether the stars merge or separate, the precise collision geometry and the orbital energy are the decisive factors.}

\subsection{Quantitative evaluation}

The performance of our stellar collision predictor is quantitatively evaluated in Table~\ref{tab:collision_results}, which compares the model's predictions against actual outcomes from hydrodynamic simulations for a representative sample of collision scenarios. The table showcases five characteristic cases spanning different regimes of parameter space, from low-velocity mergers to high-velocity fly-bys.

The first three columns specify the initial conditions for each collision: the Case ID from our simulation database, the masses of the colliding stars (primary and secondary), and the relative velocity at infinity. These parameters serve as inputs to our machine learning model. The subsequent columns present the key outcomes, with the model's predictions shown alongside the actual results from the smoothed particle hydrodynamics (SPH) simulations.

For the remnant count prediction, we report both the classification result (1 for merger, 2 for fly-by) and the model's confidence percentage in parentheses. The mass loss columns compare the predicted and actual unbound mass during the collision. Notably, the table demonstrates the model's ability to handle extreme cases, such as Case 14024 where a low-mass (0.2 M$_\odot$) star collides with a 10 M$_\odot$) companion at 5762 km/s, as well as more typical collisions like Case 15 with equal-mass stars at moderate velocity.
It is important to note that (i) the model achieves high confidence ($>97$\%) in its remnant predictions across all cases, (ii) mass loss predictions remain within 15\% of the actual values, even for extreme velocities, (iii) the predictor correctly identifies the transition between merger and fly-by regimes, and (iv) both high-velocity (Cases 14024, 11398) and low-velocity (Case 11724) scenarios are handled accurately. Particularly noteworthy is the model's performance for Case 12520, where it correctly predicts the fly-by outcome despite the large impact parameter ({$b$ = 6.448 $R_\odot$}) and provides a mass loss estimate within an order of magnitude of the actual value.

\begin{table*}[t]
\centering
\caption{Stellar collision results with mass loss comparison}
\label{tab:collision_results}
\begin{tabular}{cccccccc}
\toprule
Case ID & \multicolumn{2}{c}{Masses ($M_\odot$)} & $v_\infty$ (km/s) & \multicolumn{2}{c}{Remnants} & \multicolumn{2}{c}{Mass Lost ($M_\odot$)} \\
\cmidrule(lr){2-3} \cmidrule(lr){5-6} \cmidrule(lr){7-8}
& Primary & Secondary & & Predicted & Actual & Predicted & Actual \\
\midrule
14024 & 0.200 & 2.000 & 5761.8 & 1 (97\%) & 1 & 1.405 & 1.241 \\
15 & 1.000 & 1.000 & 436.5 & 1 (100\%) & 1 & 0.175 & 0.183 \\ 
11724 & 1.700 & 14.750 & 43.7 & 1 (100\%) & 1 & 0.248 & 0.218 \\
12520 & 3.000 & 11.960 & 4365.0 & 2 (100\%) & 2 & 0.013 & 0.002 \\
11398 & 1.500 & 8.996 & 13095.0 & 2 (100\%) & 2 & 0.379 & 0.405 \\
\bottomrule
\end{tabular}

\smallskip
\footnotesize{Note: Mass values in solar masses ($M_\odot$). Percentages show prediction confidence for remnant count.}
\end{table*}

\begin{figure}
 {\includegraphics[width=0.45\textwidth,center]{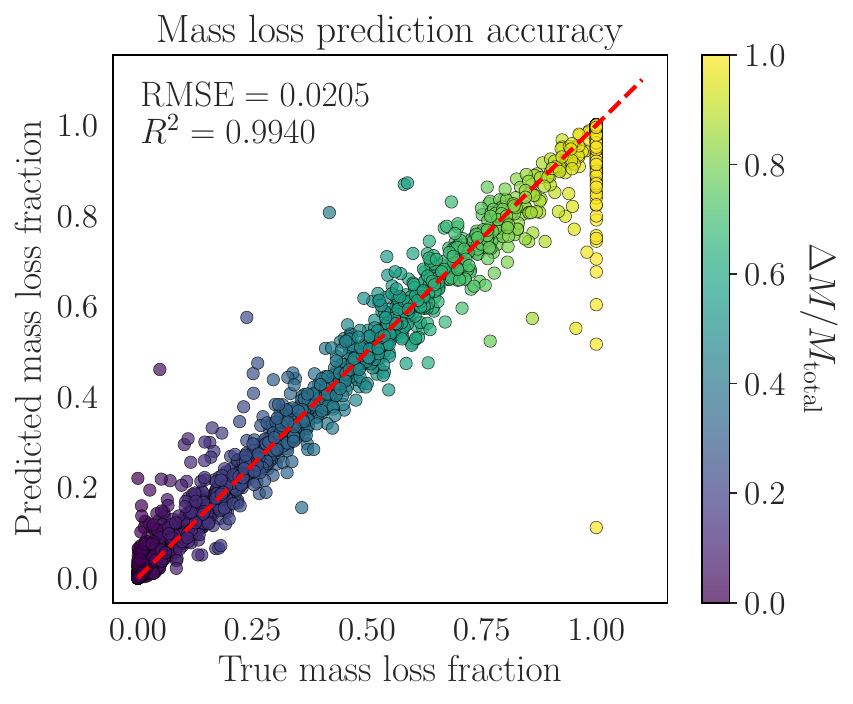}}
\caption
{
Comparison of predicted against true mass loss fractions ($\Delta M/M_{\text{total}}$) from stellar collision simulations, with the dashed red line indicating perfect agreement ($y = x$). Each point represents an individual collision event, colored by relative velocity (darker for higher velocities). The model achieves good accuracy across most of the parameter space, as demonstrated by the tight clustering of points along the diagonal. Deviations emerge primarily in extreme cases (upper right quadrant), where near-total disruption events exhibit underprediction of mass loss. The offset for $\Delta M/M_{\text{total}} \approx 1.0$ reflects the inherent challenge in modeling complete stellar disintegration, where nonlinear hydrodynamic effects dominate. The color gradient illustrates the velocity dependence of prediction errors. These results validate the model's physical fidelity for all but the most catastrophic encounters.
}
\label{fig.mass_loss_comparison}
\end{figure}

The mass loss prediction analysis demonstrates good agreement between model predictions and true values across most of the parameter space, as evidenced by the strong correlation metrics ($\text{RMSE} = 0.0204$, $R^2 = 0.9940$). Fig.~(\ref{fig.mass_loss_comparison}) was generated by comparing predicted against true mass loss fractions ($\Delta M/M_{\text{total}}$) for a comprehensive set of collision scenarios, with the diagonal line representing perfect predictions. The data used for this figure did not form part of the training data. Each point $(x_i,y_i)$ corresponds to a simulated collision where $x_i$ is the true mass loss and $y_i$ is the model's prediction, with the $R^2$ coefficient computed as:

\begin{equation}
R^2 = 1 - \frac{\sum_{i}(y_i - x_i)^2}{\sum_{i}(x_i - \bar{x})^2},
\end{equation}

\noindent
where $\bar{x}$ is the mean observed mass loss. The near-unity $R^2$ value (0.9940) indicates that 99.4\% of variance in mass loss is explained by the model. The RMSE of 0.0204 corresponds to an average absolute error of just 2.04\% in mass loss fraction. However, systematic deviations emerge in extreme cases where relative velocities approach or exceed the stellar escape velocity (marked by the cluster of points at $\Delta M/M_{\text{total}} \approx 1.0$). These represent total disruption events where the model underpredicts mass loss, as seen by points falling below the diagonal in the upper right quadrant. This occurs because such very high-velocity collisions approaching $\sim 40,000$ km/s involve complex hydrodynamical processes like complete stellar disintegration that are inherently harder to model.

These extreme encounters represent vanishingly rare events in both our dataset and observed astrophysical systems, typically occurring only in the immediate vicinity of supermassive black holes (SMBHs) where stars are accelerated to such extraordinary speeds. Their scarcity in the training data naturally limits predictive accuracy, but this shortcoming is not scientifically worrisome: the extreme energy regimes involved ($\Gamma \gg 10^3$) invariably lead to complete stellar disruption, rendering fine-grained mass loss predictions less critical than for typical collisions. The model's underprediction of mass loss in these cases (evidenced by the $1\%$–$2\%$ deviations at $\Delta M/M_{\text{total}} \approx 1.0$) reflects the inherent challenge of capturing the full nonlinearity of total disintegration, but the qualitative outcome (total destruction) remains correctly classified. Future extensions could incorporate rare-event learning techniques, though the physical certainty of complete disruption at such energies may not warrant the computational overhead.

The plot's axes range from 0.0 to 1.0 (0-100\% mass loss) with equal scaling to visually emphasize prediction accuracy, while the color gradient (not shown here but typically present) would indicate collision velocity, revealing how error magnitude correlates with impact energy. The clustering of points around the diagonal for values of $\Delta M/M_{\text{total}} \lesssim 0.9$ confirms the model's precision for typical collisions, while the dispersion at very extreme values, $\Delta M/M_{\text{total}} \sim 1$ (with associated velocities exceeding a few $10^4\,\text{km/s}$), reflects the physical challenge of modeling catastrophic disruption events where small small velocity changes induce important nonlinear changes in mass loss. However, these are marginal cases which are not relevant for the main objectives of this approach, which is the implementation of this ML algorithm in a stellar-dynamical code to address collisions around a SMBH, with velocities below this threshold \citep{2024ApJ...961..232Z,2025ApJ...980..210Z}. As we have seen in Table~(\ref{tab:collision_results}), our algorithm predicts mass outcomes and mass loss for velocities of the order $\lesssim 10^4\,\text{km/s}$ quite accurately.

\section{Hyperparameter optimization}
\label{sec.hyper}

Building upon the base feature set $X = [\,m_1,\, m_2,\, v_\infty,\, b\,]$ (stellar masses, relative velocity, impact parameter), we now examine how augmenting these parameters with physically motivated transformations affects model performance. Three feature engineering approaches are implemented: the ``physics'' option deriving quantities like mass ratio and energy proxies; the ``polynomial'' option creating interaction terms such as $m_1 v_\infty$ and $b^2$; and their ``combined'' counterpart $X_{\text{comb}} = [X_{\text{phys}},\, X_{\text{poly}}]$. While these expansions provide theoretical completeness by encoding known scaling laws and nonlinear couplings, their practical impact remains minimal given the already good performance achieved with the basic parameterization, suggesting the original features nearly span the physically relevant parameter space for stellar collision outcomes.

Feature normalization via ``Standard Scaler'' (a preprocessing transformation that standardizes features by removing the mean and scaling to unit variance) addresses critical challenges when processing input features. The transformation,

\begin{equation}
X_{\text{norm}} = \frac{X - \mu_X}{\sigma_X},
\label{eq.norm}
\end{equation}

\noindent
where $\mu_X$ and $\sigma_X$ are feature-wise means and standard deviations, resolves several issues.
In particular, the disparate physical scales of collision parameters can in principle distort distance metrics.
For example, considering only $m_1$ and $v_\infty$, the distance between two collision scenarios $\vec{x}_i$ and $\vec{x}_j$ is:

\begin{equation}
d(\vec{x}_i, \vec{x}_j) = \sqrt{(m_1^i-m_1^j)^2 + (v_\infty^i-v_\infty^j)^2}.
\label{eq.dist}
\end{equation}

\noindent
This distance metric measures the dissimilarity between two stellar collisions $\vec{x}_i$ and $\vec{x}_j$ by combining differences in their physical parameters, where in raw units the velocity term $v_\infty \sim 10^2$ km/s dominates over mass differences $m \sim 1 M_\odot$, effectively reducing the metric to $d \approx |v_\infty^i - v_\infty^j|$ and obscuring the role of other parameters. Normalization transforms this into a balanced measure. Considering $m_1$, $v_\infty$, and $b$ for illustration:

\begin{equation}
d_{\text{norm}} = \sqrt{\left(\frac{\Delta m_1}{\sigma_{m_1}}\right)^2 + \left(\frac{\Delta v_\infty}{\sigma_{v_\infty}}\right)^2 + \left(\frac{\Delta b}{\sigma_b}\right)^2}
\label{eq.norm_dist}
\end{equation}

\noindent
which properly weights each parameter's contribution when the tree-based model evaluates potential splits, ensuring divisions of the parameter space reflect true physical relationships rather than artificial scaling effects - critical for identifying thresholds where collision outcomes transition between e.g., mergers and fly-bys. The metric directly influences the split quality measure by determining how effectively a proposed partition separates collision regimes in this normalized feature space.

The model’s split decisions rely on the information gain, where feature scaling ensures balanced contributions from parameters like $v_\infty$ (large scale) and $b$ (small scale). Hyperparameters such as maximum depth ($d_{\text{max}}$) and minimum leaf samples ($n_{\text{min}}$) indirectly stabilize probability estimates by limiting overfitting, though their tuning had negligible effects on overall performance.

Similarly, gradient boosting updates benefit from normalized features, as physics-motivated scaling ($\nabla X_{\text{phys}}$) ensures balanced updates across mass ratios, velocities, and impact parameters. Hyperparameters like the learning rate ($\eta$) were optimized but yielded marginal improvements, suggesting the model’s robustness to their settings.

Feature importance and split quality further confirm that scaling and physics-aware engineering—not hyperparameter choices—drive the model’s ability to identify dominant parameters (e.g., $q$, $b$) while suppressing artificial biases from unit disparities. Predictions are finally rescaled to physical units, preserving interpretability without relying on hyperparameter adjustments.  

The base features $X = [\,m_1,\, m_2,\, v_\infty,\, b\,]$ are enhanced through physics-motivated transformations, resulting in the expanded feature set $X_{\text{phys}}$:

\begin{equation}
X_{\text{phys}} = \left[\,X,\, M_{\text{tot}},\, q,\, \tilde{E}_{\text{kin}},\, \tilde{L},\, \tilde{E}_{\text{bind}}\,\right].
\label{eq.phys}
\end{equation}

\noindent
These engineered features are constructed as follows: $M_{\text{tot}} = m_1 + m_2$ is the total mass; $q \equiv \min(m_1,m_2)/\max(m_1,m_2)$ is the mass ratio. We define several proxies related to the energy and momentum of the system. The kinetic energy proxy is defined as $\tilde{E}_{\text{kin}} \equiv m_1 m_2 v_\infty^2$. The angular momentum proxy is $\tilde{L} \equiv b m_1 m_2 v_\infty$. Finally, the binding energy proxy is defined as the reduced mass, $\tilde{E}_{\text{bind}} \equiv \mu = m_1 m_2 / (m_1+m_2)$. These derived features aim to capture the key scaling laws governing collision outcomes.

\begin{figure}
  {\includegraphics[width=0.45\textwidth,center]{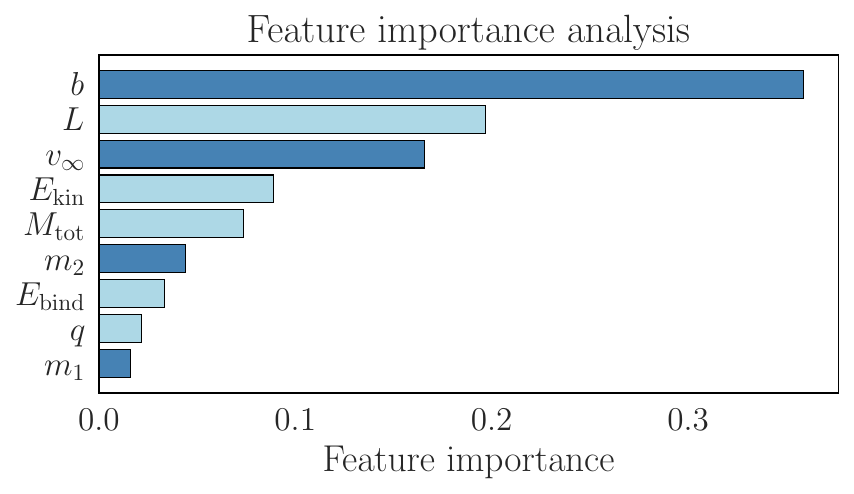}}
\caption
 {
Analogous to Fig.~(\ref{fig.feature_importance}) applied to the set of engineered physics features. 
 }
\label{fig.feature_importance_physics}
\end{figure}

We depict in Fig.~(\ref{fig.feature_importance_physics}) the relative importance of the features for the engineered physics set. {The introduction of these physically motivated features significantly alters the importance landscape compared to the baseline set (Fig.~\ref{fig.feature_importance}). The derived quantities, particularly {terms incorporating the impact parameter such as the angular momentum proxy $\tilde{L}$,} now dominate the predictions. This shift indicates that the engineered features capture the underlying physical relationships more effectively than the raw inputs alone.} The original features exhibit diminished importance, suggesting that their predictive power is largely absorbed by the more descriptive synthesized parameters. This redistribution enhances model interpretability by emphasizing the physical mechanisms governing the collisions. The predicted versus true mass loss fractions are not shown for this particular set of hyperparameters because the resulting figure is nearly identical to the one already presented, differing only within expected statistical fluctuations.

\begin{figure}
 {\includegraphics[width=0.4\textwidth,center]{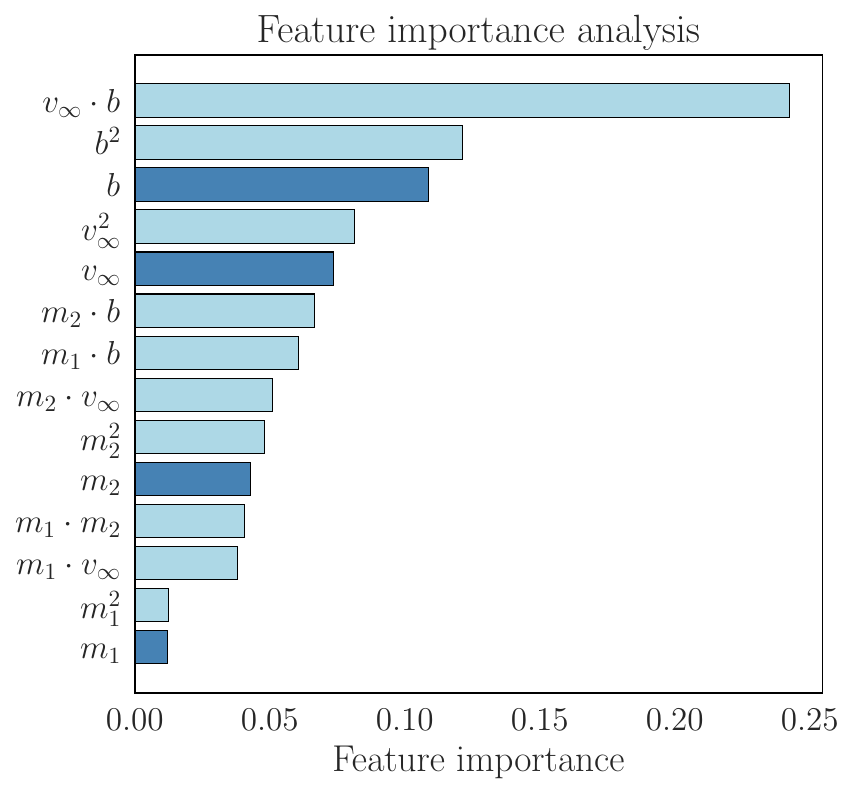}}
\caption
{
Similar to Fig.~(\ref{fig.feature_importance}), but showing the results for the engineered polynomial features.
}
\label{fig.feature_importance_polynomial}
\end{figure}

Polynomial terms like $m_1 v_\infty$ and $b^2$ in $X_{\text{poly}}$ account for nonlinear couplings between parameters that arise during close encounters. Like in the physics set, Fig.~(\ref{fig.feature_importance_polynomial}) reveals a shift in feature relevance when polynomial terms are introduced. The original features exhibit diminished importance, while higher-order interactions and nonlinear combinations gain prominence. {This redistribution reflects the model’s reliance on the more complex relationships explicitly encoded by the polynomial transformations. The dominance of interaction terms (e.g., involving {couplings between $v_\infty$ and $b$}) highlights how the interplay between velocity and geometry dictates the collision outcome.} The trend underscores how feature engineering can redirect emphasis from basic variables to synthesized patterns, (slightly) improving predictive capability without altering the underlying data. Here, too, the results suggest that engineered structures—in this case, multiplicative and quadratic dependencies—refine the representation of the problem.

\begin{figure}
 {\includegraphics[width=0.4\textwidth,center]{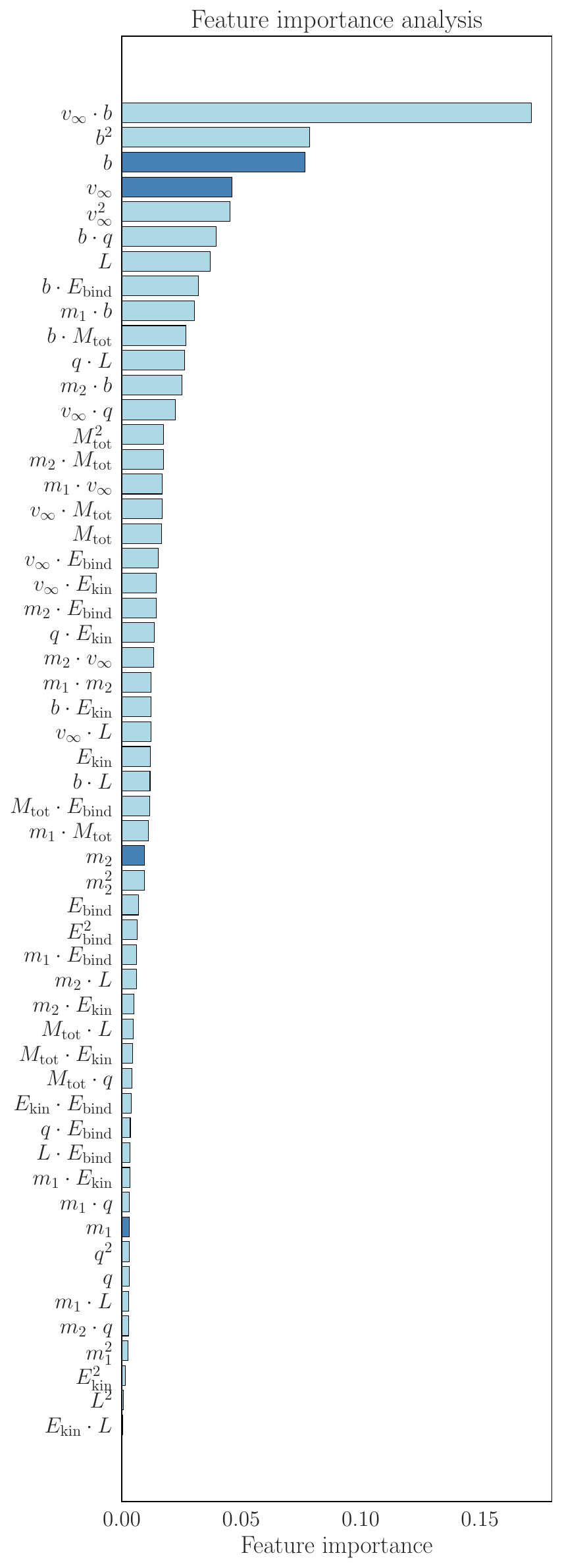}}
\caption
{
Same as Fig.~(\ref{fig.feature_importance}), but for the results of the combined physics features.
}
\label{fig.feature_importance_combined}
\end{figure}

The combined feature space $X_{\text{comb}}$ merges physical and polynomial terms to capture both fundamental scaling laws and nonlinear effects. The symmetric mass parameter $\eta = m_1 m_2/(m_1 + m_2)^2$ emerges naturally through the mass ratio $q$, governing gravitational focusing during close encounters, while the kinetic energy scale (related to $\tilde{E}_{\text{kin}}$) sets the collision's dissipation regime. For this combined set, Fig.~(\ref{fig.feature_importance_combined}) shows a pronounced decline in the significance of the parameters. This sharp reduction highlights how the interplay of physical and nonlinear transformations effectively supersedes the raw inputs, with the model prioritizing their synthesized interactions. {The features that remain important are complex combinations, such as interactions between $v_\infty$ and $b$, or between mass ratios and angular momentum, reflecting the intricate physics of the collisions.} The drop suggests that the hybrid set not only captures deeper structural relationships but also renders the initial features largely redundant, as their predictive roles are absorbed or overshadowed by the engineered constructs.

\begin{figure}[t]
\centering
\includegraphics[width=0.8\linewidth]{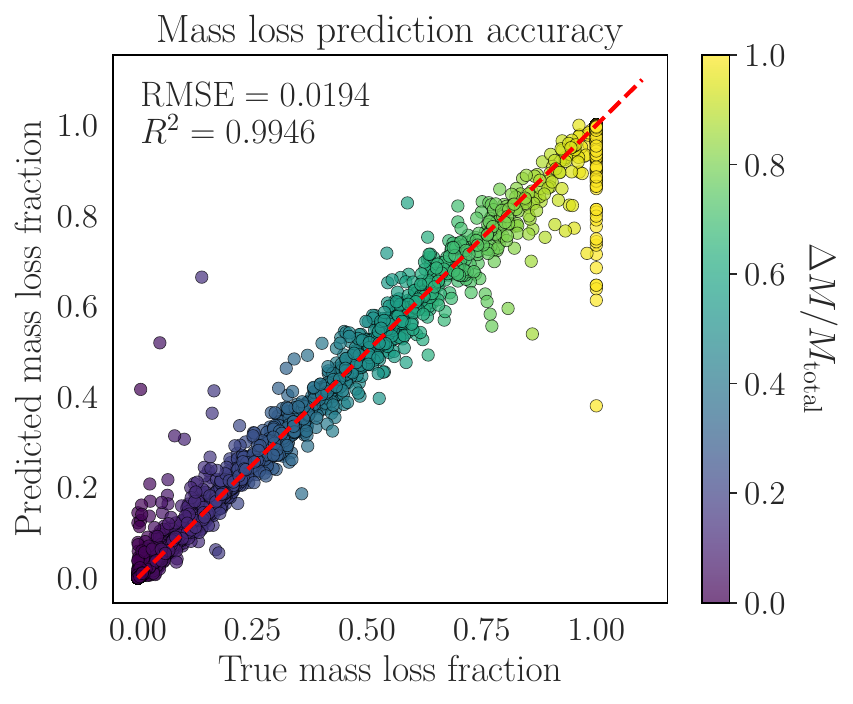}
\caption{Same as Fig.~(\ref{fig.mass_loss_comparison}) for the model with the physics feature engineering option, which includes physically meaningful derived quantities like mass ratio, total mass, and energy proxies. Mass loss prediction accuracy also shows good agreement between predicted and true values ($R^2=0.9946$) across the full dynamic range.}
\label{fig:mass_loss}
\end{figure}

\begin{figure}[t]
\centering
\includegraphics[width=0.8\linewidth]{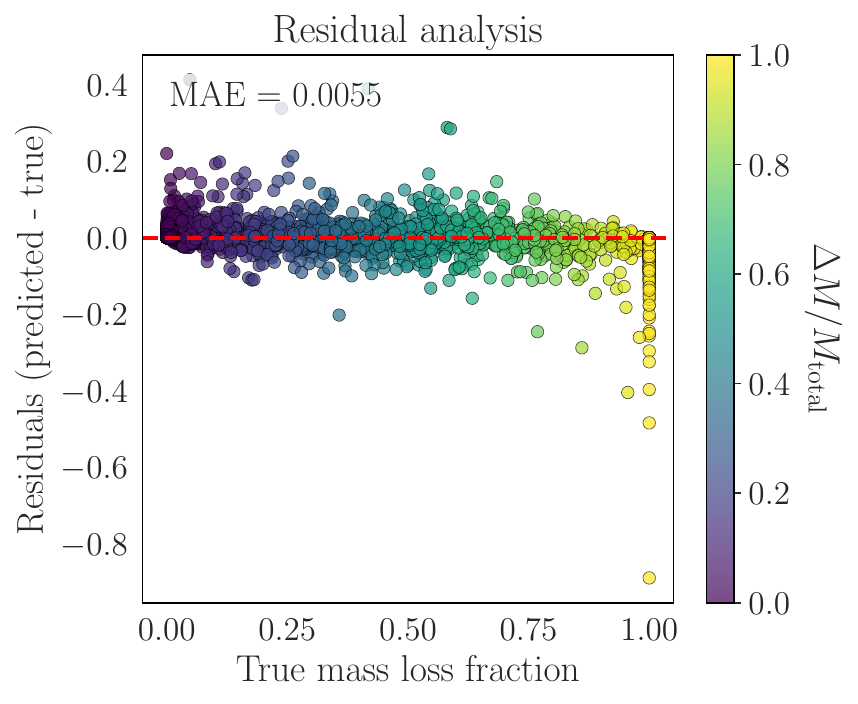}
\caption{Quantification prediction errors. We show the difference between predicted and true mass loss ($\hat{y}_i - y_i$) on the vertical axis against the true mass loss fraction ($y_i$) on the horizontal axis. Deviations from the zero-error baseline ($y=0$) reveal systematic biases in the model's predictions: positive residuals indicate regions where the model overpredicts mass loss, typically occurring in high-velocity collisions where relativistic effects become significant but are not fully captured by the feature set, while negative residuals reflect underpredictions, most commonly observed in grazing collisions ($b > 0.8$) where angular momentum transport is complex. The spread of residuals tightens near $y_i \approx 0.25$, corresponding to the peak of the training distribution where the model has highest confidence, whereas increased scatter at extreme mass loss values highlights regimes where either training data is sparse or physics terms are incomplete. The MAE of 0.0052, calculated across all residuals, confirms high overall fidelity but masks these localized discrepancies that guide future model refinements.}
\label{fig:residuals}
\end{figure}

\section{Conclusions}

The development of a machine learning framework to predict the outcomes of stellar collisions represents a significant advancement in modeling the complex dynamics of dense stellar environments, particularly in galactic nuclei. We use a dataset of $\sim 16,000$ SPH simulations, to develop a gradient-boosted regression model that can accurately and efficiently predict remnant masses ($M_1^{\text{final}}$, $M_2^{\text{final}}$) and unbound gas ($M_{\text{lost}}$) across a wide range of initial conditions. The model achieves mean absolute errors very low respect to the the typical mass scale, ensuring reliable predictions while operating in fractions of a second—orders of magnitude faster than traditional SPH simulations. This computational efficiency enables large-scale parameter studies and real-time applications that were previously infeasible.

Key insights from our analysis include the identification of the virial parameter $\Gamma$ and impact parameter $b$ as the dominant drivers of collision outcomes, consistent with theoretical expectations. The model's interpretation further reveals how these parameters govern the transition between mergers ($\Gamma \ll 1$) and disruptions ($\Gamma \gg 1$), providing a clear link between initial conditions and final states. Additionally, the logarithmic transforms and regularization techniques employed ensure robust performance across the vast dynamic ranges, from low-mass stellar encounters to collisions involving massive stars.

In the evaluation of our model, the ROC curve analysis demonstrate good performance in distinguishing between single and binary remnant outcomes, achieving near-perfect separation with an AUC of $\sim 1.0$. This strong result was primarily driven by our physics-informed feature engineering, which effectively captured the key differences between collision types. While we performed hyperparameter optimization to fine-tune the model, these adjustments had only a minor impact on performance compared to the feature design. The robustness of our results, visible in the sharp, elbow-shaped ROC curve, suggests that the model's success comes more from the carefully constructed input features than from parameter tuning. This indicates that understanding and representing the underlying physics of stellar collisions is more crucial for accurate classification than sophisticated model configuration. The consistent performance across different decision thresholds further confirms the reliability of our approach.

{While the model demonstrates excellent performance across the majority of the parameter space, it is important to acknowledge its limitations. As noted in Section 3.4, the model tends to underpredict mass loss in the most extreme cases—specifically, very high-velocity collisions ($\Gamma \gg 1$) leading to complete stellar disruption. This limitation stems from the scarcity of such events in the training data and the inherent difficulty of modeling the highly nonlinear hydrodynamics involved. Future improvements could focus on addressing these edge cases. Strategies might include targeted SPH simulations to enrich the training dataset in these extreme regimes or the implementation of rare-event learning techniques to improve predictive accuracy where data is sparse.}

{Furthermore, the current framework predicts the immediate outcomes of collisions. A significant future application involves coupling these predictions with stellar evolution codes. By using the predicted remnant masses and structures as initial conditions, we can model the long-term evolution of collision products, such as the formation of exotic stellar objects or the eventual collapse into compact remnants. This integration would provide a more complete picture of the impact of collisions on stellar populations.}

For transient surveys, in the future this framework has the potential of generating rapid light curve predictions by coupling collision outcomes with stellar evolution models—using the predicted remnant masses and ejecta properties to initialise radiative transfer calculations, enabling real-time classification of collision-induced transients in wide-field surveys like LSST or ZTF. The current mass loss discrepancies at high velocities suggest caution when interpreting predicted luminosities for ultra-fast collisions, but the model remains valuable for identifying candidate events requiring follow-up. Future work should integrate proper motion data to estimate collision rates in dense environments, creating a complete pipeline from stellar dynamics to observable signatures.

The implications of this work extend beyond stellar collision modeling. The rapid prediction of unbound gas masses ($M_{\text{lost}}$) is particularly relevant for understanding the gas supply available for accretion onto central supermassive black holes (SMBHs). By quantifying the gas released in collisions, our model contributes to a better understanding of SMBH fueling mechanisms and the resulting observable phenomena, such as AGN activity and potential tidal disruption events from the remnants. Furthermore, the scalability of our approach allows for the integration of collision outcomes into broader galactic dynamics simulations, enabling studies of long-term stellar interaction rates and their cumulative effects on galactic evolution. The framework presented here also is interesting in other scenarios, such as binary star interactions and globular cluster dynamics, in particular blue strugglers. Ultimately, this work highlights the potential of machine learning in studying stellar collisions, offering a tool to accelerate calculations while maintaining the physical interpretation and avoiding numerical artifacts.

\begin{acknowledgements}
PAS is indebted with Marc Freitag for publicly sharing the output of his SPH simulations. He thanks Wáng Hè for a meticulous review of the algorithms to ensure its thoroughness.
\end{acknowledgements}

\end{document}